# Fluid flow anlaysis in a rough fracture (type II) using complex networks and lattice Boltzmann method

H. Ghaffari
*Department of Civil Engineering, Lassonde Institute, University of Toronto, Toronto, ON, Canada*

A. Nabovati
*Department of Mechanical & Industrial Engineering, University of Toronto, Toronto, ON, Canada*

M. Sharifzadeh
*Faculty of Mining Engineering, Tehran Polytechnic, Tehran, Iran*

R. P. Young
*Department of Civil Engineering, Lassonde Institute, University of Toronto, Toronto, ON, Canada*

ABSTRACT
Complexity of fluid flow in a rough fracture is induced by the complex configurations of opening areas between the fracture planes. In this study, we model fluid flow in an evolvable real rock joint structure, which under certain normal load is sheared. In an experimental study, information regarding about apertures of the rock joint during consecutive 20 mm displacements and fluid flow (permeability) in different pressure heads have been recorded by a scanner laser. Our aim in this study is to simulate the fluid flow in the mentioned complex geometries using the lattice Boltzmann method (LBM), while the characteristics of the aperture field will be compared with the modeled fluid flow permeability To characterize the aperture, we use a new concept in the graph theory, namely: complex networks and motif analysis of the corresponding networks. In this approach, the similar aperture profile along the fluid flow direction is mapped in to a network space. The modeled permeability using the LBM shows good correlation with the experimental measured values. Furthermore, the two main characters of the obtained networks, i.e., characteristic length and number of edges show the same evolutionary trend with the modeled permeability values. Analysis of motifs through the obtained networks showed the most transient sub-graphs are much more frequent in residual stages. This coincides with nearly stable fluid flow and high permeability values.

RÉSUMÉ
La complexité de l'écoulement du fluide dans une fracture rugueuse est induite par la configuration complexe de l'ouverture des zones parmi les plans de fracture. Dans cette étude, nous écoulement du modèle de fluide dans un joint de vrai rock évolutif qui sous une charge normale de certains est cisaillé. Information en ce qui concerne d'ouvertures de la roche commune pendant 20 mm déplacements et des flux de fluide (perméabilité) dans la tête des pressions différentes ont été enregistrées par balayage laser. Notre objectif dans cette étude est de simuler l'écoulement de fluide dans les géométries complexes mentionnés en 2D LBM tandis que les caractéristiques du champ d'ouverture seront comparés avec les attributs modèle d'écoulement de fluide. Pour la caractérisation des champs dans les ouvertures, nous utilisons une nouvelle théorie de la théorie des graphes, à savoir, les réseaux complexes et l'analyse des motifs des réseaux correspondants. De cette façon, le profil d'ouverture similaires le long de la direction de l'écoulement du fluide est mappé dans un espace réseau. La perméabilité modélisés à l'aide LBM montre une bonne corrélation avec les valeurs expérimentales mesurées. En outre, les deux personnages principaux des réseaux obtenus, à savoir, la longueur caractéristique et le nombre de bords montrent la même tendance évolutive avec les valeurs de perméabilité modélisés. L'analyse des motifs à travers les réseaux obtenus ont montré le plus transitoire sous-graphes sont beaucoup plus fréquents dans les stades résiduelle. Cela coïncide avec écoulement du fluide à peu près stable et les valeurs à haute perméabilité.

## 1 INTRODUCTION

Analysis of fluid flow in a rough fracture has allocated a notable research scope in geoseicence. Cracks, fractures, and joints are the main passages of the fluid flow in rocks (geomaterials). The collection of the opening and void elements induces their own complexities. Complexity investigation is noticed by analyzing and inspecting the behavior of the system against the forced (assumed) attributes. For instance, the attributes of joint's (fractures/cracks) swarms are the reactions to mechanical, hydraulic, chemical, thermal, electrical or magnetic (or a combination of these) forces.

To capture more complexity through the system, rather than assumption of mean field or simplified rules, single element behavior has to be considered, e.g., a single fracture (or rock joint). Particularly, analysis of hydraulic property (like advection and diffusion) within a fracture, which is under several stresses, is much more relevant in practical engineering works. Considering developments of apertures (openings) with shear displacements and recording the hydraulic conductivity, which is measured during the laboratory coupled shear-flow tests, have been the subject of a few studies in recent years [1-3]. In addition to macro scale interwoven nature of fractures, the micro-scale properties of the fracture planes (halves) show and insert their own complexities to the problem. Thus, topological complexities bring new additional coupling in time-space of the coupled attributes based problems. Flow of information (fluid or energy) through such interwoven structures is generally characterized by permeability, conductivity or transmissivity. Consequently, measuring and estimation of flow quality and complexity index has always been a major research subject.

Among different approaches in modeling the fluid flows within the complex geometries – from numerical methods perspective and especially to estimate the permeability- the lattice Boltzmann method (LBM) has attracted a lot of attention in the last two decades [9,10]. Madadi et al. [4] studied fluid flow in a two-dimensional model of a fracture with rough, self-affine internal surfaces using the LBM for predicting flow properties. The fracture model consisted of two parallel flat plates on which two rough, self-affine surfaces, characterized by a roughness exponent, H, were superimposed.

Auradou et al. [5] investigated the induced permeability anisotropy by the shear displacement using the LBM, while the results were compared with the experimental results. Eker and Akin [6] performed numerical analysis on synthetic rock fracture using the LBM considering the fractal dimension of fractures..The estimated permeability values using the LBM were less than the cubic law. Kim et al. [7] presented numerical computations for single phase flow through three-dimensional digitized rock fractures that came from profiled elevations taken on tensile induced fractures in Harcourt granite under varied simulated confining pressures. Numerical predictions of fracture permeability were compared with the laboratory measurements performed on the same fractures. They concluded that the use of the finite difference LBM allows computation on non-uniform grid spacing, which enables accurate resolution across the aperture width without extensive refinement in the other two directions.

In this study, we use a two-dimensional single-relaxation time LBM to model fluid flow in a real rock joint, where the rock joint under constant normal load is sheared. The topology of fracture surfaces (top and bottom) is used to construct pore spaces (aperture field). We will be using two different approaches for modeling fluid flow. In the first situation, the aperture fields are considered as the pore space. In the second case, the mean aperture of profiles (in the direction of the flow - parallel to the shear direction) is used to build a channel-like profile where the bottom surface for simplicity is assumed to be a flat plane. Comparing the results obtained from these two approaches, we can study the accuracy of the mean profiles (slice method) in estimating the permeability. In order to study the connection between aperture's structural complexity, fluid flow and mechanical deformation of the rock joint, a graph theory on the aperture profiles is used. Characterization of opening spaces with the properties of the constructed networks will discriminate the role of the contact and non-contact areas to conduct information flow. The organization of this study is as follows. In the second part the LBM and networks method are explained. The next section briefly demonstrates the experiment process. The last part presents the results from the LBM and complex network, which is followed by a brief discussion and conclusion.

## 2 MATERIALS AND METHODS

In this section, we briefly summarize the 2D LBM with nine discrete velocities (D2Q9). Also, the procedure of building networks over aperture field is discussed.

### 2.1 LBM modeling of 2D Incompressible Fluid Flow

Chronologically, the LBM is the result of the efforts to improve the Lattice Gas Cellular Automata (LGCA) [8-10]. The LBM predicts the distribution function of fictitious fluid particles on fixed lattice sites on discrete time steps in discrete directions. Two main steps in the LBM algorithm are the streaming and collision. In the streaming step, particles move to the nearest neighbor along their direction-wise discretized velocities. The streaming process is followed by collision step, where the particles relax towards a local equilibrium distribution function.

In the single relaxation time LBM the collision operator is simplified to include only one relaxation time for all the modes. In the D2Q9 LBM, the fluid particles at each node are allowed to move to their eight nearest neighbours with eight different velocities, $\mathbf{e_i}$. The ninth particle is at rest and does not move. The fluid density and macroscopic velocities are calculated by properly integrating the particle distribution functions on each node. The evolution of distribution function in single-relaxation time LBM [11] is given by:

$$f_i(\mathbf{x}+\mathbf{e_i}\Delta t, t+\Delta t) = f_i(\mathbf{x},t) - \frac{\Delta t}{\tau}[f_i(\mathbf{x},t) - f^{eq}{}_i(\mathbf{x},t)] \quad (1)$$

where for any lattice node, $x + e_i\Delta t$ is its nearest node along the direction *i*; $\tau$ is the relaxation time; $f^{eq}{}_i$ is the equilibrium distribution function in direction *i*. The macroscopic fluid variables, density $\rho$ and velocity can be obtained from the moments of the distribution functions as follows:

$$\rho = \sum f_i ; \rho\mathbf{v} = \sum f_i \mathbf{e_i} \quad (2)$$

while the fluid pressure field p is determined by the equation of state of an ideal gas:

$$p = C_s{}^2 \rho \quad (3)$$

where $C_s$ is the fluid speed of sound and for D2Q9 lattice is equal to $\frac{1}{\sqrt{3}}$.

The relaxation time is characterizing the time-scale behaviour of fluid particle collisions and determines the lattice fluid viscosity:

$$\nu = \frac{1}{3}(\tau - \frac{1}{2}) \quad (4)$$

Once the macroscopic velocity field is determined, the permeability of the medium under study can be predicted using Darcy's law [12]:

$$\langle \mathbf{v} \rangle = -\frac{\mathbf{K}}{\mu}.\nabla p \quad (5)$$

where $\langle \mathbf{v} \rangle$, $\mathbf{K}$, $\nabla p$ and $\mu$ are the volume averaged flow velocity, permeability tensor, pressure gradient vector and the dynamic viscosity of the fluid, respectively. The fluid flow is driven by pressure (or by some external force) on the boundaries. To extract the correct permeability values, the system should be in the steady state. The computation is in the steady state when the relative change of the average velocity is less than a tolerance variable (here $10^{-9}$). More details about the different numerical and technical aspects of the LBM can be found elsewhere [8-12].

### 2.2 Complex Networks and Sub-graphs on Void Spaces

A network consists of nodes and edges that are connecting them [13]. To set up a network on apertures (openings) of a fracture under a certain value of normal stress, a network is constructed on the enclosed interval among two surfaces, i.e., complex aperture networks. The main point in the selection of each space is to explore the explicit or implicit hidden relations among different distributed elements of a system. To set up a network, we consider each profile of aperture as a node. If we assume the lower surface is fixed then the motion of upper surface causes elimination of some nodes. Then, we focus on the intersecting nodes. To make an edge between two nodes, we use correlation measurement over the aperture profiles. For each pair of profiles) $V_i$ and $V_j$ containing N elements (pixels), the correlation coefficient can be written as:

$$C_{ij} = \frac{\sum_{k=1}^{N} [V_i(k) - \prec V_i \succ] . [V_j(k) - \prec V_j \succ]}{\sqrt{\sum_{k=1}^{N} [V_i(k) - \prec V_i \succ]^2} . \sqrt{\sum_{k=1}^{N} [V_j(k) - \prec V_j \succ]^2}} \qquad (6)$$

where $\prec V_i \succ = \frac{\sum_{k=1}^{N} V_i(k)}{N}$. Obviously, it should be noted that $C_{ij}$ is restricted to $-1 \le C_{ij} \le 1$, where $C_{ij}=1$, 0, and -1 are related to perfect correlations, no correlations and perfect anti-correlations, respectively. We set $C_{ij} \ge \xi = 0.2 C_{ij}^{\max}$.

Considering this definition, we are filtering uncorrelated profiles over the metric space. Total number of edges and characteristic length are two main properties of the graphs. The average (characteristic) path length *L* is the mean length of the shortest paths connecting any two nodes on the graph. The shortest path between a pair (i, j) of nodes in a network can be assumed as their geodesic distance, $g_{ij}$, with a mean geodesic distance *L* given as below ([14], [15]):

$$L = \frac{2}{N(N-1)} \sum_{i<j} g_{ij}, \qquad (7)$$

where $g_{ij}$ is the geodesic distance (shortest distance) between node *i* and *j*, and *N* is the number of nodes. We will use a well known algorithm in finding the shortest paths presented by Dijkstra [16]. For a given network with N nodes, the degree of the edge and Laplacian of the connectivity matrix are defined by [17]:

$$k_i = \sum_{j=1}^{N} A_{ij} ; L_{ij} = A_{ij} - k_i \delta_{ij} \qquad (8)$$

where $k$, $A_{ij}$, $L_{ij}$ are the degree of $i^{th}$ node, elements of a symmetric adjacency matrix and the network Laplacian matrix, respectively. The eigenvalues $\Lambda_\alpha$ are given by $\sum_{j=1}^{N} L_{ij}\phi_j^\alpha = \Lambda_\alpha \phi_i^\alpha$ in which $\phi_i^\alpha$ is the $i^{th}$ eigenvector of the Laplacian matrix ($\alpha = 1,...,N$). Using this definition, all of the eigenvalues are non-positive values. The inverse participation ratio as a criterion to localization of eigenvectors is defined by [18]:

$$P(\boldsymbol{\varphi}^\alpha) = \frac{\sum_i \phi_i^\alpha}{(\sum_i (\phi_i^\alpha)^2)^2} ; \alpha = 1,...,N \qquad (9)$$

The maximum value of *P* is one which shows the vector has only one non-zero component and then all of the weight of the vector is spread through one node. Higher value of *P* corresponds to a more localized vector.

From the other point of view, the granulations of internal structures (local building blocks) through the obtained networks are presented by sub-graphs and motifs. The sub-graphs are the nodes within the network with the special shape(s) of connectivity together. The relative abundance of sub-graphs has been shown is an index to functionality of networks with respect to information processing. Furthermore, they are in correlation with global characteristics of the networks [19-22]. Network motifs introduced by Milo et al. [19] are particular sub-graphs representing patterns of local interconnections between the nodes in the network. A motif is a sub-graph that appears more than a certain number, other criteria can be found in elsewhere in the literatures [references]. A motif of size *k* (containing *k* nodes) is called a *k*-motif (or generally sub-graph). We employ the above mentioned approaches over the networks of aperture profiles.

### 2.3 Summary of Laboratory Tests

This section summarizes the results of several laboratory tests on a rock joint. The joint geometry consisting of two joint surfaces and the aperture (opening or void space in 3D) between these two surfaces were measured. The shear and flow tests were performed later on. The rock was granite with a unit weight of 25.9 kN/m3 and uniaxial compressive strength of 172 MPa. An artificial rock joint was made at mid height of the specimen by splitting and using special joint creating apparatus, which has two horizontal jacks and a vertical jack [23-24]. The sides of the joint are cut down after creating the joint. The final size of the sample is 180 mm in length, 100 mm in width and 80 mm in height. A virtual mesh having a square element size of 0.2 mm was spread on each surface and the height at each position was measured by a laser scanner. In the experiments, a special hydraulic testing unit is used to allow linear flow (parallel to the shear direction) intermittently with the shear displacements of the rock joint. The relation between the hydraulic conductivity and the joint aperture is given by Darcy's law (Eq. 6). The details of the procedure about the reconstruction of the aperture fields can be found in [23-24].

## 3 RESULTS AND DISCUSSION

To model the fluid flow in the joint, the structured of the flow passage is mapped in to the binary values, zero for solid nodes and one for fluid nodes. LBM simulations of fluid flow in the flow passages corresponding to 4 different shear displacements are presented.

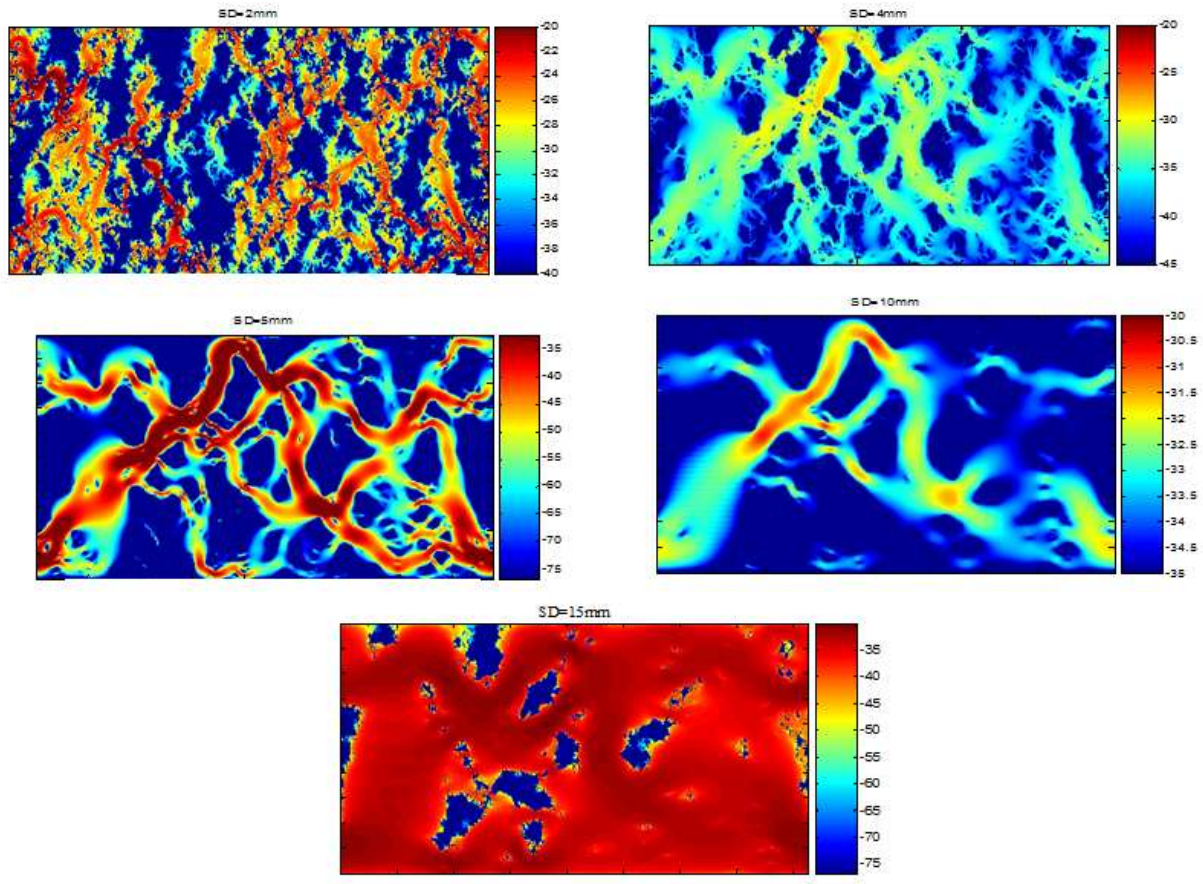

Figure 1. Velocity field obtained from LBM simulation over the aperture field and through 4 shear displacements (SD): 2, 4, 5, 10 and 15 mm (Logarithmic scale of velocity) –To discriminate channelization effect, we have used interactive color bars (legend). The resolution factor is 1 then each pixel is 0.2 mm.

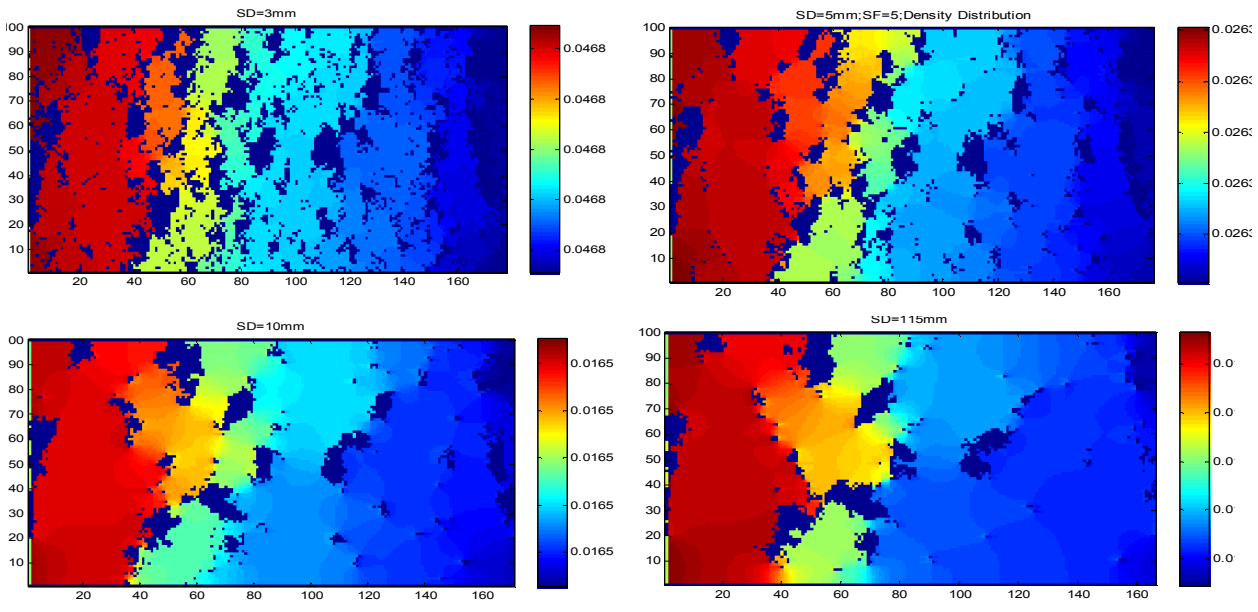

Figure 2. Density Distribution (proportional to pressure distribution) in logarithmic scale through successive shear displacement: a) 2 mm, b) 5 mm, c) 10 mm and d) 15 mm (resolution factor 5; each pixel is equal to 1 mm)..

For the first case, using full resolution (~800×500 computational nodes), the velocity field patterns through successive shear displacements has been shown in Fig. 1. As it can be followed for all of the cases, the channelization effect is clear and dominant: main portion of the fluid is flowing through pathways with minimum resistance. This channelization effect is the result of the contact areas, where the contact areas generally are decreasing with mean aperture of the fracture. In other words, increasing the mean aperture field is roughly equivalent to decreasing contact areas.

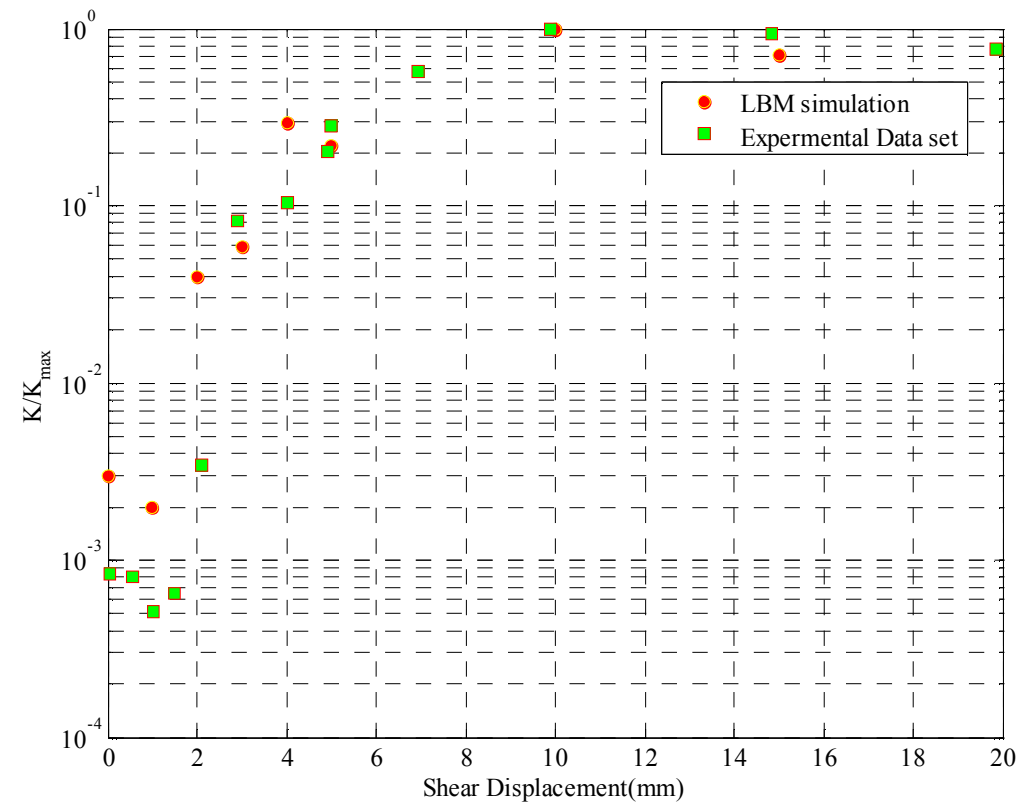

Figure 3. Normalized (to maximum value) permeability values obtained from LBM simulation on aperture fields (circle) and experimental results (rectangle) versus successive shear displacements of the fracture.

The fluctuations in contact area, integrated with aperture variation and thus mechanical volume change (called dilitancy), is correlated with the development of shear forces. At initial displacements (<2 mm), due to the interlocking of the asperities of the joint, the opening spaces are decreasing that will directly affect the magnitude contact areas (increasing the contact areas). This phenomenon is restricting the information flow, i.e., dropping permeability (see figure 3 around 0-1 mm). A comparison of the normalized numerically predicted permeability values against the experimentally measured permeability values has been shown in Fig. 5. As it can be seen, there is a good agreement between the numerical and experimental data points. However, numerical estimation of permeability in the interlocking stage (between 0 mm to near 2 mm of shear displacement) using the LBM predicts values larger than the experimentally measured values. This could be justified by considering that the variation of permeability for the fractures with large roughness is more rapid and more prone to geometry change rather than the fractures with small values of the roughness [4]. It was observed that for small aperture fields the time to reach a steady state is much higher than that required for larger opening spaces. This can bring a new idea with respect to information flow in most interwoven porous structures. In other words, trapping of information (flow) in interlocking stage of the evolvable fracture (while the frictional forces are following an incremental trend) is much higher and induces more complexity in finding flow pathways through the fracture.

In order to perform a more comprehensive analysis on permeability patterns and possible effects of the orientation of contact areas, the fluid flow is modeled across to shear direction where the shear direction is in X direction. As it has been illustrated in figure 4, the main path of the flow is demonstrating the channelization effect. However, the fluid flow in this case is better facilitated in comparison to the simulations where the mean flow direction was parallel to the shear displacement (see Fig. 5). In other words, the shear forces induce anisotropy in the predicted permeability. It is noteworthy that with approaching to the last shear displacements, the induced anisotropy will be limited to a nearly a constant value.

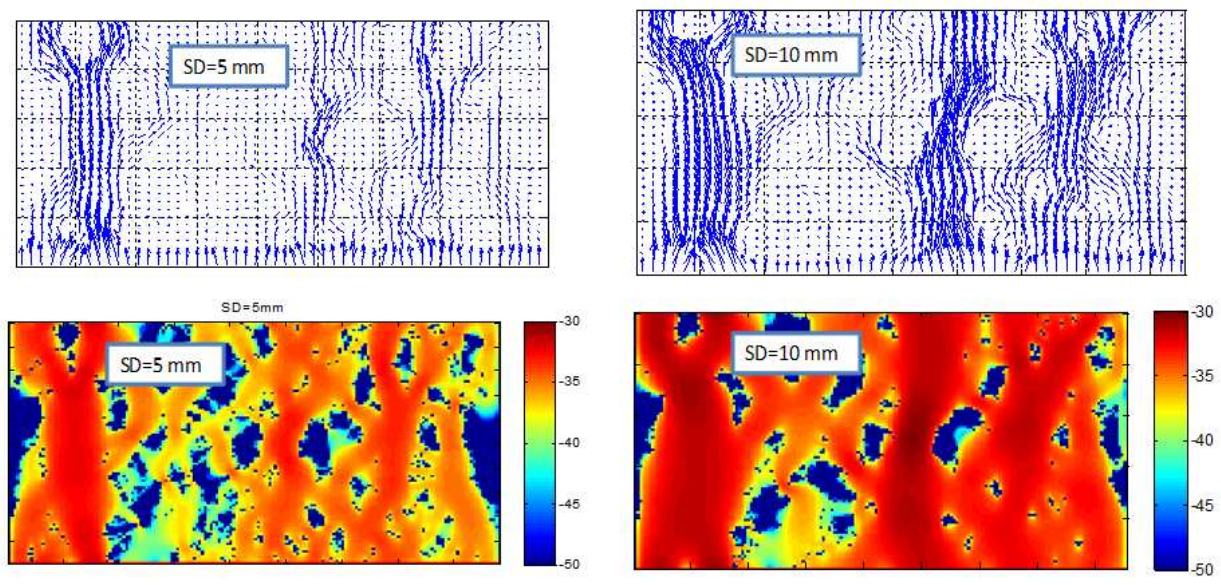

Figure 4. The results of fluid flow modeling along Y-direction (perpendicular to shear direction) on two cases: shear displacement (SD) =5 and 10 mm. The resolution factor is 5 and velocity fields are in a logarithmic scale.

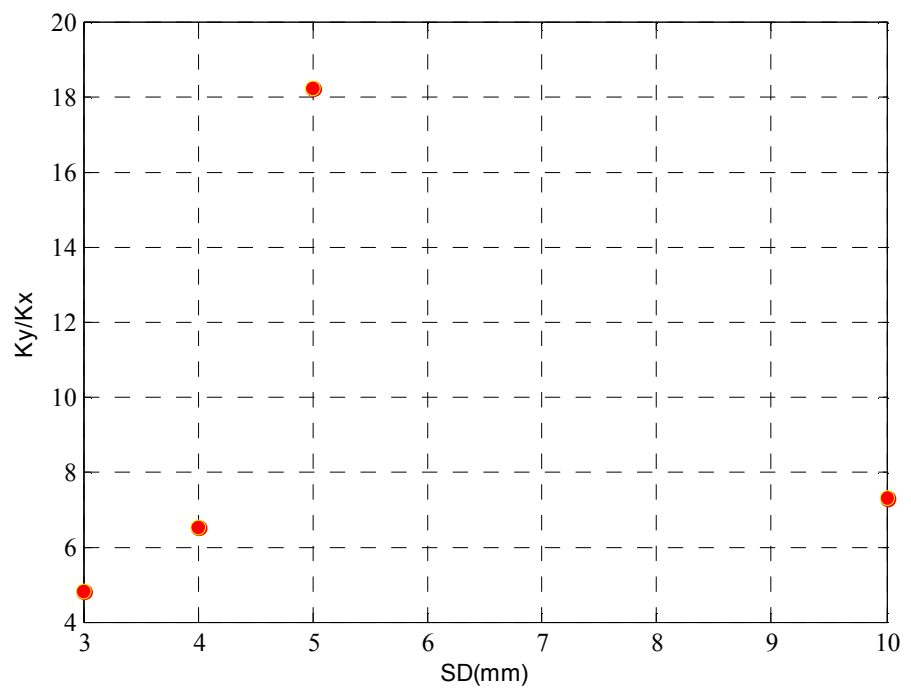

Figure 5. Anisotropy of permeability during 10 mm successive shear displacements

To end up our analysis around permeability analysis, we map the parallel (with shear direction) aperture profiles into the network space with the method described in section 2.2. We will compare the properties of the obtained networks with the fluid flow patterns, which are produced by LBM and experimental results (Figure. 6a). The characteristic length of networks exhibits a rapid drop after passing the interlocking step (Figure. 6b - coincides with nearly after peak point of shear stress-displacement). The same temporal evolutionary trend in inverse of mean geodesic length of the parallel aperture profiles with the measured hydraulic conductivity gives the idea of formation huge clusters over the parallel networks (Figure. 6c). This shows how the correlated profiles are propagating and diffusing (virtually) through fracture while the shear displacement is progressing. This is absolutely in agreement with increment of permeability and easy flow information. Then, the characteristic length of the obtained network can be scaled with the simulated permeability.

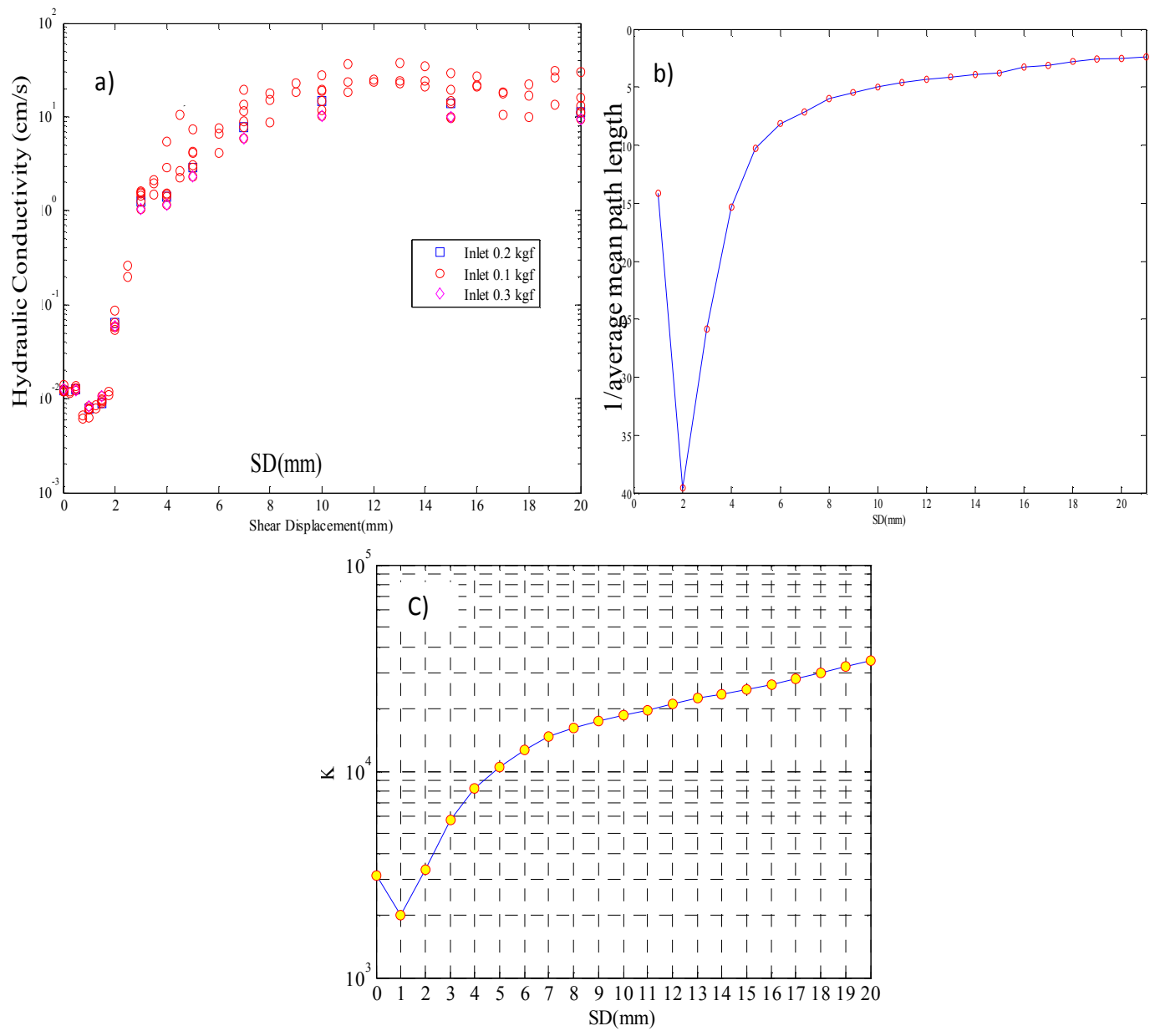

Figure 6. Comparison of (a) hydraulic conductivity (measured through experimental laboratory tests) and (b) the inverse of the mean geodesic length of the parallel aperture networks (flipping Y-axis) and (c) evolution of edges degree with displacements in semi-logarithmic scale.

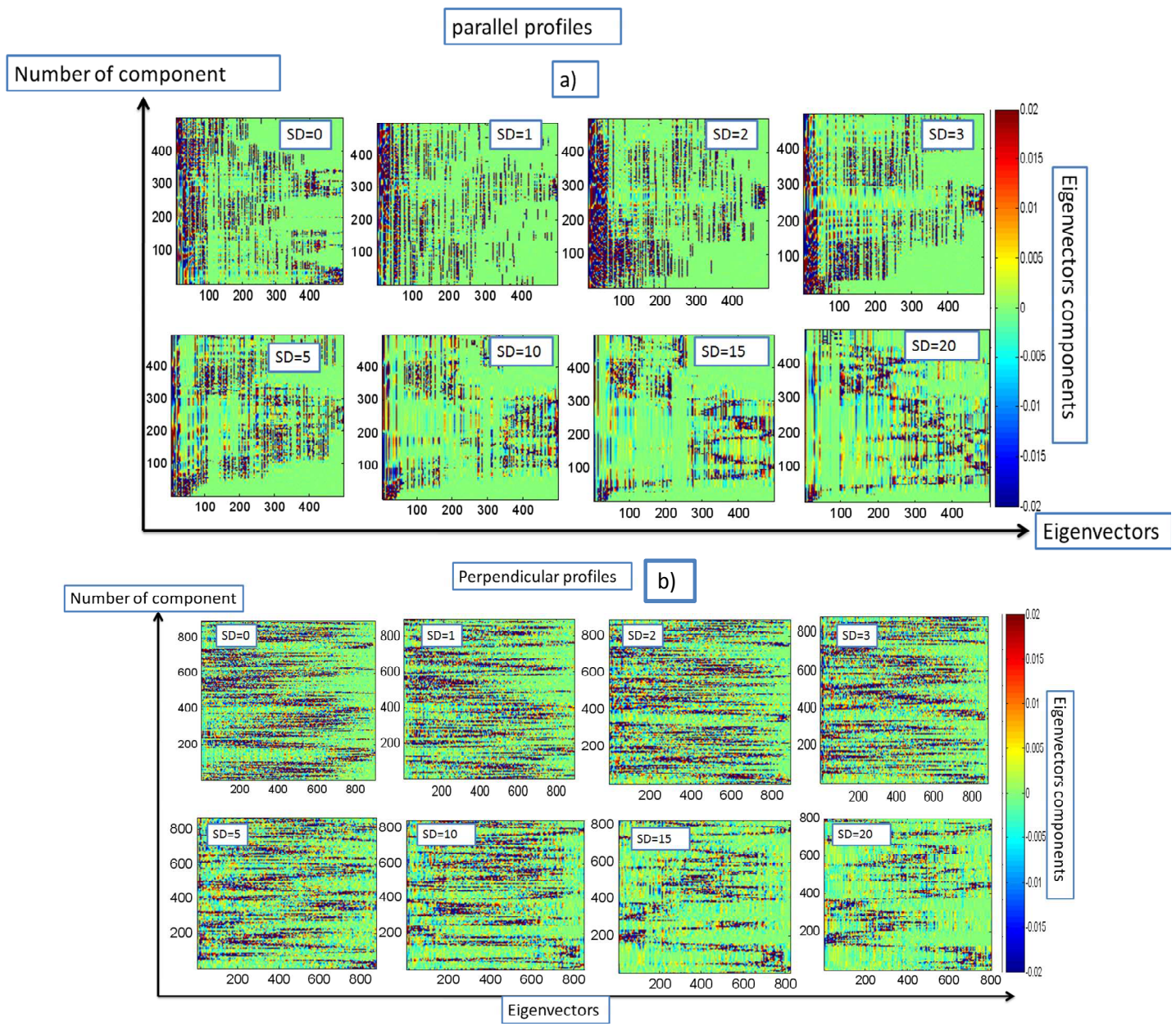

Figure 7. Visualization of eigenvectors components in a) parallel profiles-Networks, and b) perpendicular profile-networks. The horizontal axes are showing the eigenvector and the vertical axes are the component (Number of nodes) of the eigenvectors.

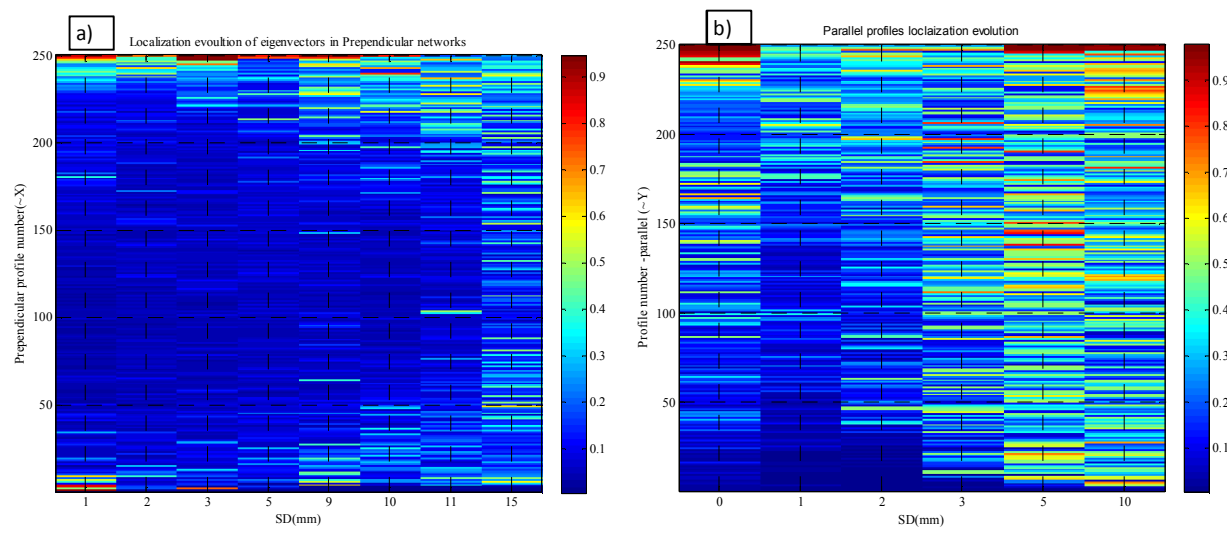

Figure 8. Evolution of the inverse participation factor for profiles in X and Y directions: a) perpendicular profiles are localized from boundaries to inside profiles. b) Parallel profiles after interlocking of asperities (SD~1mm) are nearly following random localization.

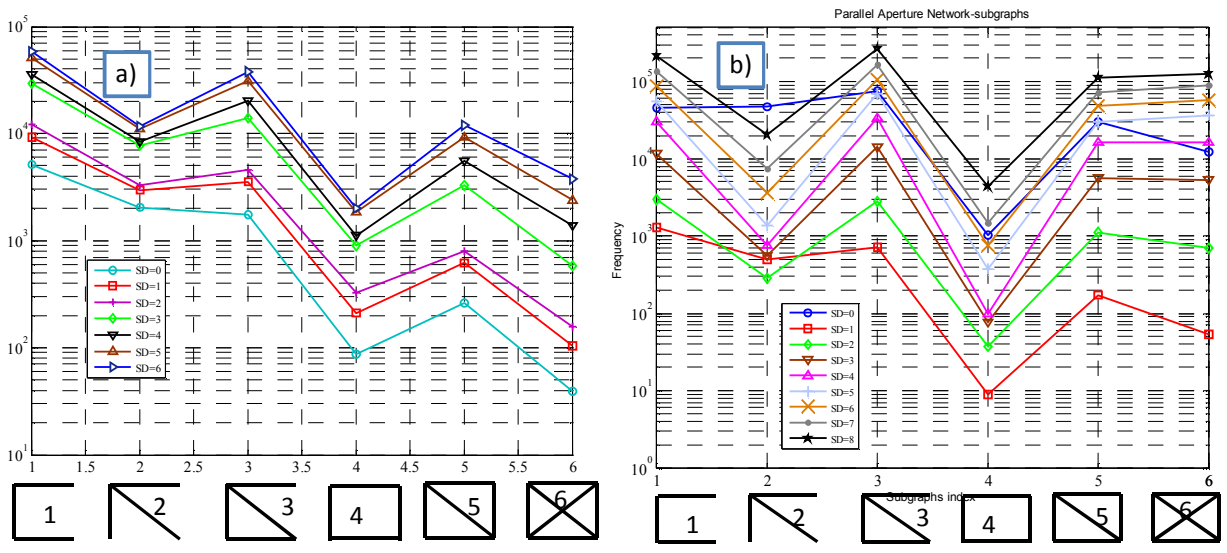

Figure 9. Evolution of 4-points sub-graphs for profiles in X and Y directions: a) perpendicular profiles, b) Parallel profiles.

In Figs. 7a and 7b, we present the evolution of eigenvectors patterns for parallel and perpendicular networks. The components with very high and low values are presenting the patterns of vectors. As one can follow, the differentiated components reveal certain patterns- either for parallel or perpendicular networks- in each time evolution. Regular patterns (last shear displacements) are coinciding with more localization of vectors (see Fig. 8). Appearance of ordered patterns (then less entropy) is much clear in parallel networks rather than perpendicular ones. In other words, the patterns in perpendicular profiles (where mostly controlling the frictional output of the system) are much complex than parallel profiles. Evaluation of localization of eigenvectors for perpendicular networks (Figure 8) shows localization is propagating from boundaries to inside profiles while in parallel networks, after interlocking of asperities (SD~1mm), is nearly following random localization patterns. For both cases, the boundary profiles are much more localized than interior profiles. The localization of eigenvectors is much more remarkable in parallel profiles rather than perpendicular ones. The abundance of the localized elements for the last time steps and especially after the Coulomb threshold level is following the same trend of permeability evolution. Apparently, generating a network with localized eigenvectors includes the easy flow of information. For mechanical interpretation, one can assume the patterns of aperture are the potential of the fracture to rupture and driving the fracture length. In other words, before the slip, the trapping of energy through contact zones are equivalent with the minimum localization of eigenvectors of the corresponding networks; while after the slip the frequency of the localized patterns of the eigenvectors are higher. Analysis of 4-points sub-graphs over both perpendicular and parallel networks (with reducing of the resolution) are shown in Fig. 9. A jumping of the abundance of sub-graphs patterns in SD=2mm to 3mm (Fig. 9 a) is in agreement with the mechanical deformation and dilatancy of the fracture. Also, dropping of all sub-graphs from SD=0mm to SD=1 mm is the index of interlocking and dramatic drop of the permeability (Fig. 9 b). The in common property for both perpendicular and parallel sub-graphs is the main evolutionary trend of sub-graphs. Especially, after the slip, the most transient sub-graphs (index 6) are showing faster growth and percolation over the both networks. The same evolutionary trend is predicted for directional profiles (neither parallel nor perpendicular) and is showing the same mechanism in functionality of networks (the same mechanism in evolution of fracture). The rapid increment of index 6 especially for parallel networks which changing the main trend of the frequency-index space, is in absolute agreement with easy fluid flow in residual stages of the sheared fracture. The low value of index 2, 4 is showing the localization of flow, i.e., channelization effect. Because the appearance of index 2 and 4 is much more relevant to heterogeneity of flow and is in contrast with channelization effect. Then, much more stable flow patterns are accompanied with the most transient sub-graphs like index 6.

## 4 CONCLUSION

The study of fracture and fracture networks are the main subject of diverse research area. Especially, estimation of fracture conductivity and permeability is a major challenge in several fields of engineering. In this study we used the LBM and networks theory to model the flow pattern through an evolvable rough fracture. With having the fracture surfaces topology and the measured permeability in the direction of shear displacement, we confirmed the validity of our model results. The channelization effect was dominant both for parallel and perpendicular flow. The anisotropy in permeability was observed in the LBM predicted results, while the permeability in perpendicular direction was higher than parallel one and fluctuating during the shearing process. Furthermore, we characterized the aperture profiles with network attributes, where among them the characteristic length was showing the same trend as the permeability evolution. This shows how correlations of complex void spaces are controlling the fluid flow path. Analysis of motifs through the obtained networks showed the most transient sub-graphs are much more frequent in residual stages. This coincides with nearly stable fluid flow and high permeability values.